# A Comparative Study on Monte Carlo Simulations of Electron Emission from Liquid Water

[1]Mehnaz, [1]L.H. Yang, [2]Y.B. Zou, [3]B. Da, [4]S.F. Mao[§] and [1]Z.J. Ding[*]

[1]Hefei National Laboratory for Physical Sciences at Microscale and Department of Physics, University of Science and Technology of China, Hefei, Anhui 230026, P.R. China;

[2]School of Physics & Electronic Engineering, Xinjiang Normal University, Urumchi, Xinjiang 830054, P.R. China

[3]Center for Materials Research by Information Integration (CMI2), Research and Services Division of Materials Data and Integrated System (MaDIS), National Institute for Materials Science, 1-2-1 Sengen, Tsukuba, Ibaraki 305-0047, Japan

[4]Department of Engineering and Applied Physics, University of Science and Technology of China, Hefei, Anhui 230026, People's Republic of China

[§]e-mail: sfmao@ustc.edu.cn

[*]e-mail: zjding@ustc.edu.cn

**Abstract:** Liquid water has been proved to be an excellent medium for specimen structure imaging by a scanning electron microscope. Knowledge of electron-water interaction physics and particularly the secondary electron yield is essential to the interpretation of the imaging contrast. However, very little is known up to now experimentally on the low energy electron interaction with liquid water because of certain practical limitations. It is then important to gain some useful information about electron emission from water by a Monte Carlo (MC) simulation technique that can numerically model electron transport trajectories in water. In this study, we have performed MC simulations of electron emission from liquid water in the primary energy range of 50 eV-30 keV by using two different codes, i.e. a classical MC (CMC) code developed in our laboratory and the Geant4-DNA (G4DNA) code. The calculated secondary electron yield and electron backscattering coefficient are compared with experimental results wherever applicable to verify the validity of physical models for the electron-water interaction. The secondary electron yield vs. primary energy curves calculated by the two codes present the same generic curve shape as that of metals but in rather different absolute values. G4DNA yields the underestimated absolute values due to the application of one step thermalization model by setting a cutoff energy at 7.4 eV so that the low energy losses due to phonon excitations are omitted. Our CMC calculation of secondary electron yield is closer to the experimental data and the energy distribution is reasonable. It is concluded that a full dielectric function data at low energy loss values below 7.4 eV shall be employed in G4DNA model for the modeling of low energy electrons.

metals but in rather different absolute values. G4DNA yields the underestimated absolute values due to the application of one step thermalization model by setting a cutoff energy at 7.4 eV so that the low energy losses due to phonon excitations are omitted. Our CMC calculation of secondary electron yield is closer to the experimental data and the energy distribution is reasonable. It is concluded that a full dielectric function data at low energy loss values below 7.4 eV shall be employed in G4DNA model for the modeling of low energy electrons.

# I. Introduction

Water, ubiquitous liquid in nature and as an essential life material is involved in radiation interactions in the fields of biophysics, environmental radiation protection, radiation dosimetry, space radiation physics etc. Moreover, it is also concerned with scanning electron microscopy (SEM) imaging of biomaterials [1-3]. But due to difficulties to perform scattering experiments with liquid water in vacuum, there has been only a very limited amount of experimental work performed on electron-water (liquid state) interactions. With development of technology it is now possible to perform such an experiment with an environmental SEM [2-3]. However, the measurement with these techniques has been limited to high energy electrons (> 5 keV), and experiments with low energy electrons are still quite restricted.

On the other hand, Monte Carlo (MC) computer simulation methods have extensively been applied in many fields to explore the electron-matter interaction mechanisms [4-5] and can therefore, provide a useful means to quench the increased thirst for the interaction mechanisms and for obtaining the required quantitative data. By a MC simulation technique, one calculates statistically moving electron trajectories in a sample which are formed by successive scattering events for incident electrons and signal electrons. The most important issue for an accurate MC simulation is the physical modeling of electron elastic and inelastic scattering in a medium. There are many different approaches available now for calculating elastic and inelastic scattering cross-sections under different approximations, and hence many corresponding MC codes have been developed based on these approaches. In our previous studies we have developed a MC simulation model based on a dielectric functional theory to treat electron inelastic scattering in a conductive solid [6-7]; this model has been successfully applied to the study of surface electron spectroscopic and SEM signals [8-12]. Besides there are also several other MC models and codes aimed mainly at study of electron interaction with conductive solids [13], and some MC codes have been developed to model electron-water interactions [14-19] and among them Geant4-DNA (G4DNA) [17-18] has received more attention.

In this work we therefore aim to study secondary electron emission and electron backscattering from water by using two MC simulation models and codes, i.e. our CMC (classical trajectory MC) code and G4DNA code. The simulation results on secondary electron yield (SEY) and electron backscattering coefficient in the primary energy range of 50 eV-30 keV are compared with the available experimental data to validate the MC simulation modeling.

## II.  CMC Physical Modeling

While most of the present MC simulations are based on the classical trajectory concept of electron motion by random sampling of moving path and scattering fate, the present CMC physical modeling employ quantum mechanical formulations for electron elastic and inelastic scattering cross-sections. The effectiveness of the model for the simulation of cascade secondary electron generation in conductive solids has been confirmed by comparing the simulated energy spectra and the SEYs with the experimental data [7-9]. This direct MC simulation of cascade production and emission of secondary electrons can provide quantitative physical quantities about secondary electrons without need of any fitting parameters and hence an illuminating insight into secondary electron emission phenomenon. In this work, we directly extend the calculation model from the conductive solid to liquid water. Although an improved simulation model is demonstrated to be quite useful for investigation of charging phenomenon in an insulating solid [20], the charging effect is negligible for liquid water.

1. Electron Elastic Scattering

At low energy region elastic scattering is the main interaction process for electrons. But there lacks experimental data for electron elastic scattering in liquid water at low energies. For the description of electron elastic scattering with an atomic-nuclei, which is responsible for the change of direction of electron velocity vector, the relativistic quantum mechanical formulation via Mott's differential cross-section [21] has been employed:

$$\frac{d\sigma_e}{d\Omega} = |f(\theta)|^2 + |g(\theta)|^2, \tag{1}$$

where $\theta$ is the scattering angle. The scattering amplitudes,

$$f(\theta) = \frac{1}{2ik} \sum_{l=0}^{\infty} \left\{ (l+1)\left(e^{2i\delta_l^+} - 1\right) + l\left(e^{2i\delta_l^-} - 1\right) \right\} P_l(\cos\theta); \tag{2}$$

$$g(\theta) = \frac{1}{2ik} \sum_{l=1}^{\infty} \left\{ -e^{2i\delta_l^+} + e^{2i\delta_l^-} \right\} P_l^1(\cos\theta), \tag{3}$$

are calculated by a partial wave expansion method [22], where $\delta_l^+$ and $\delta_l^-$ are spin up and spin down phase shifts of the $l$ th partial wave, respectively; $P_l(\cos\theta)$ and $P_l^1(\cos\theta)$ are the Legendre and the first order associated Legendre functions, respectively. In this work, we adopted the static-field approximation with Dirac-Fock atomic electron densities [23] and Fermi-distribution nuclear densities. The Furness-McCarthy exchange potential [24] is added to account for exchange effects while polarization and absorption are ignored. The calculation of Mott cross-section has been performed by using the computer code ELSEPA [25]. The same scattering potential was used to simulate reflection electron energy loss spectroscopy spectra of metals [26]. This scattering potential is considered to be much more accurate than other potential, such as Thomas-Fermi-Dirac atomic potential. It is well known that the Mott cross-section exceeds the Rutherford formula based on classical mechanics at low electron energies and particularly for heavy elements due to spin-orbit coupling during electron collision with the nuclei [27]. For the calculation to liquid water, only the molecular density and chemical composition of $H_2O$ have been considered in deriving molecular scattering cross-section from individual atomic components by ignoring the molecular structure.

2. Electron Inelastic Scattering

For modeling electron inelastic scattering in a dielectric medium, the Penn's formalism based on a dielectric functional approach has been employed. The differential inverse electron inelastic mean free path (DIIMFP) for an electron with kinetic energy $E$ in the first Born approximation is given by

$$\frac{d^2\lambda_{in}^{-1}}{dqd\omega} = \frac{\hbar}{\pi a_0 E} \text{Im}\left\{\frac{-1}{\varepsilon(q,\omega)}\right\} \frac{1}{q}, \tag{4}$$

where $\lambda_{in}$ denotes the electron inelastic mean free path (IMFP) in a condensed matter, $a_0$ is Bohr radius, and $\hbar\omega$ and $\hbar q$ are the energy loss and the momentum transfer of the electron, respectively. $\varepsilon(q,\omega)$ is the dielectric function of the medium and $\text{Im}\{-1/\varepsilon(q,\omega)\}$ is called energy loss function (ELF) which completely determines the probability of an inelastic scattering event, the energy loss distribution, and the scattering angular distribution of the electron.

For modeling ELF, the full Penn's algorithm [28,29] has been employed here,

$$\text{Im}\left\{\frac{-1}{\varepsilon(q,\omega)}\right\} = \int_0^\infty g(\omega_p) \text{Im}\left\{\frac{-1}{\varepsilon_L(q,\omega;\omega_p)}\right\} d\omega_p, \tag{5}$$

where $\varepsilon_L(q,\omega;\omega_p)$ is the Lindhard dielectric function for a free electron gas with plasmon energy $\hbar\omega_p$, which is a function of local electron density. $g(\omega)$ is the expansion coefficient and is related to the optical ELF (i.e. in the optical limit $q \to 0$) by $g(\omega) = (2/\pi\omega)\text{Im}\{-1/\varepsilon(\omega)\}$. This algorithm can yield a better agreement with experimental results for SEY and energy distribution curve than the previous modeling by considering a single-pole approximation [6-10] where $\text{Im}\{-1/\varepsilon_L(q,\omega;\omega_p)\}$ is simplified as a Dirac $\delta$-function along the plasmon dispersion line [30].

The Lindhard ELF is made of two individual parts, i.e. the single electron excitation part $\text{Im}\{-1/\varepsilon(q,\omega)\}_e$ and plasmon excitation part $\text{Im}\{-1/\varepsilon(q,\omega)\}_{pl}$, as:

$$\text{Im}\left\{\frac{-1}{\varepsilon(q,\omega)}\right\} = \text{Im}\left\{\frac{-1}{\varepsilon(q,\omega)}\right\}_e + \text{Im}\left\{\frac{-1}{\varepsilon(q,\omega)}\right\}_{pl}. \tag{6}$$

Hence, by the full Penn's algorithm the low energy secondary electron generation from Fermi sea has two distinct mechanisms, either through single electron excitation or via a plasmon decay [12,29]. By the property of the Lindhard ELF, the single electron excitation occurs if $q^-(\omega;\omega_p) < q < q^+(\omega;\omega_p)$ and plasmon excitation occurs if $q < q^-(\omega;\omega_p)$, where $q^-$ and $q^+$ are, respectively, the left and right boundaries of the area for the non-vanishing imaginary part of the Lindhard dielectric function. While for $\hbar\omega > \hbar\omega_p$ when $E > \hbar\omega_p$, the secondary electron generation can only be the single electron excitation and, particularly, by the inner-shell ionization when the kinetic energy exceeds the binding energy $E_B$ of atomic electrons, as $\text{Im}\{-1/\varepsilon(q,\omega)\}_{pl} \approx 0$ in this case. The inclusion of single electron excitation part reasonably solves a problem in the single-pole approximation, i.e. lacking the energy loss channel when $E < \hbar\omega_p$ for a free-electron metal.

Once the ELF is determined, the energy loss distribution and the electron IMFP, $\lambda_{in}$, can be respectively evaluated by the integrations,

$$\frac{d\lambda_{in}^{-1}}{d\omega} = \int_{q_-}^{q_+} \frac{d^2 \lambda_{in}^{-1}}{dq d\omega} dq, \tag{8}$$

and

$$\lambda_{in}^{-1} = \int_0^{E-E_F} \frac{d\lambda_{in}^{-1}}{d\omega} d\omega, \tag{9}$$

where the integration limits, $\hbar q_{\pm} = \sqrt{2m}\left(\sqrt{E} \pm \sqrt{E-\hbar\omega}\right)$, are the largest and the smallest momentum transfers kinematically allowed for a given energy $E$ and energy loss $\hbar\omega$. The restriction to $\hbar\omega \leq E - E_F$ is due to the Pauli exclusion principle that an electron cannot fall into the Fermi sea which is already occupied by electrons. The energy loss distribution is used for sampling a specific energy loss value in an inelastic scattering event, while IMFP with total elastic scattering cross-section determines the transport mean free path used for sampling a flight step length in a MC simulation.

As for optical data, experimental optical constants are used to derive the complex dielectric constant, $\varepsilon(\omega) = \varepsilon_1(\omega) + i\varepsilon_2(\omega)$, and optical ELF, $\text{Im}\{-1/\varepsilon(\omega)\}$ as input to the simulation. Use of experimental data allows an accurate description of phonon excitation and electronic excitation processes responsible for electron energy loss in a realistic material. Fig. 1 shows the experimental data of optical ELF for liquid water as a function of energy loss [31-33]. There is no obvious difference on ELF between liquid, hexagonal ice and amorphous ice, but liquid phase differs largely with gas phase in the mid-energy loss range of 10-80 eV. The energy loss and the rate due to the excitation of rotational and translational phonon modes are much lower than the electronic excitation considered here for secondary electron generation [34]; the energy loss due to phonon excitation below 1 eV cannot contribute to emitted secondary electron signals but plays an important role to slow down electrons including the cascade secondary electrons inside water. ELF indicates that the dominant contribution to the energy loss and the associated secondary electron generation is the electronic excitation around several tens eV, which is mostly associated molecular orbitals of water and does not have obvious characteristic excitation edges. An obvious sharp ionization edge around $E_B = 532$ eV is due to K-electrons of oxygen atoms, which plays an important role to the electron stopping power.

3. Secondary Electron Generation

According to this full Penn algorithm there is no difference on the treatment of energy loss in an inelastic scattering event due to valence electron excitation and the inner shell ionization, but production of secondary electron by these two distinct loss mechanisms differs. Here only the K shell of oxygen is considered as an inner shell. In an inelastic scattering event an energy loss $\hbar\omega$ is firstly sampled with a uniform random number from Eq. (8). If $\hbar\omega < E_B$, then a secondary electron is assumed to be excited from the Fermi sea by transferring $\hbar\omega$ from the moving electron to a valence electron of energy with the excitation probability being proportional to a joint density of states of free electrons $\sqrt{E_0(E_0+E_F)}$, where $E_0 < E_F$ is the energy of the valence electron. For the mechanism of secondary electron generation via bulk plasmon decay [35], we assumed that the energy loss for bulk plasmon excitation is then immediately transferred to a secondary electron. An excited secondary electron may undergo the similar electron inelastic scattering to produce multiple lower energy secondary electrons and cause a cascade production. For secondary electron penetration from the surface barrier into vacuum to become emitted signals, a quantum description of the transmission probability is adopted [7]. The transmission function can be written as:

$$T(E,\beta) = \begin{cases} \dfrac{4\sqrt{1-U_0/E\cos^2\beta}}{\left[1+\sqrt{1-U_0/E\cos^2\beta}\right]^2}, & \text{if } E\cos^2\beta > U_0, \\ 0, & \text{otherwise,} \end{cases} \qquad (10)$$

Where $\beta$ is the ejection angle of electrons measured from surface normal direction, and the inner potential is taken as the electron affinity, i.e. $U_0 = E_A$.

By using this direct MC simulation of secondary electron production, the SEY value, $\delta$, can be evaluated by taking the ratio of the number of emitted true secondary electrons ($E < 50$ eV) to the number of simulated incident electrons at a given primary beam energy $E_p$.

## III.  G4DNA Physical Modeling

G4DNA is the low energy extension of Geant4 [14-16, 19], which is an open source MC simulation toolkit based on object orientated programming rules, coded with the C++ language, created and developed at CERN. Liquid water has been chosen as an interacting medium for the application of G4DNA for its importance to radiobiological simulations. McNamara et al. have used Geant4 to investigate the low energy secondary electron track structures produced by x-ray and proton beams in liquid water [36]. G4DNA employs physics models for electron interaction processes by including elastic scattering, electronic excitation and ionization in the kinetic energy range of 7.4 eV- 1 MeV [17-18, 37-46]. Like CMC all interactions are treated as discrete processes, i.e. an electron does not lose its kinetic energy along the flight step length and interaction happens always at the flight step terminal. A moving electron can produce a secondary electron once the energy loss in an inelastic scattering is higher than the binding energy of the target electron.

1. Electron Elastic Scattering

In G4DNA, elastic scattering cross-sections are computed either by a screened Rutherford model or by a partial wave model. Screened Rutherford model has the advantage that differential and total elastic scattering cross-sections are analytic, which is considered most suitable for fast MC simulations [43-44]. Here we used the improved screened Rutherford model implemented in G4DNA which has a better agreement with the more elaborate partial wave model.

The approach applied in Geant4 consists of extending the following elastic scattering models for the free atoms (and molecules) to the liquid phase. For electron energies above 200 eV, the screened Rutherford differential scattering cross-section is adopted [40,44]:

$$\frac{d\sigma_e}{d\Omega} = \frac{Z(Z+1)e^4}{4E^2(1+2\alpha-\cos\theta)^2}, \quad (11)$$

where $Z$ is the atomic number, $\alpha$ is screening parameter given by Uehara et al. [43-45] based on the available scattering data:

$$\alpha = \frac{1.198KZ^{2/3}}{E(2+E/mc^2)}, \quad (12)$$

where $K = 1.75 \times 10^{-5}$ and $mc^2 = 511003.4$ eV are constants. The total cross-section is then calculated by integrating the differential cross-section as,

$$\sigma_e = 2\pi \int_0^\pi \frac{d\sigma_e}{d\Omega} \sin\theta d\theta = \frac{\pi Z(Z+1)e^4}{4E^2\alpha(\alpha+1)}. \quad (13)$$

At energies below 200 eV, Brenner and Zaider's empirical formula is used which fits the available experimental elastic scattering angular distribution data of water vapor [40-41,47];

$$\frac{d\sigma_e}{d\Omega} \propto \frac{1}{(1+2\gamma-\cos\theta)^2} + \frac{\beta}{(1+2\delta-\cos\theta)^2}, \quad (14)$$

where the fitting parameters are expressed as [43]:

$$\gamma(E) = \begin{cases} \exp\left(\sum_{n=0}^{5}\gamma_n E^n\right), & 0.35 \text{ eV} \leq E \leq 10 \text{ eV}; \\ \exp\left(\sum_{n=0}^{4}\gamma_{n+6} E^n\right), & 10 \text{ eV} < E \leq 100 \text{ eV}; \\ \sum_{n=0}^{2}\gamma_{n+11} E^n, & 100 \text{ eV} < E \leq 200 \text{ eV}, \end{cases} \quad (15)$$

$$\beta(E) = \exp\left(\sum_{n=0}^{4}\beta_n E^n\right), \quad (16)$$

$$\delta(E) = \exp\left(\sum_{n=0}^{4}\delta_n E^n\right). \quad (17)$$

2. Electron Inelastic Scattering

To calculate electron inelastic scattering cross-section a dielectric formulation is used with a correction at low energy. The calculation of ionization and excitation cross-sections [37-44] for incident electron with energy <10 keV is based on the Emfietzoglou semi-empirical inelastic model [42,45] of the dielectric function with Heller's optical data [48]. It makes use of (a) the dielectric formalism for the valence shells responsible for condensed-phase effects and (b) the binary-encounter-approximation for the K-shell of oxygen atoms. According to this model, the electronic structure of liquid water can be represented by four "outer" ionization shells, one "inner" ionization shell (the K-shell of oxygen), and five discrete excitation levels.

A. Valence band

The optical ELF is partitioned into ionizations and excitations as follows:

$$\text{Im}\{\varepsilon(\omega)\} = \sum_{n=1}^{4}\text{Im}\{\varepsilon_n(\omega)\} + \sum_{k=1}^{5}\text{Im}\{\varepsilon_k(\omega)\}$$
$$= \sum_{n=1}^{4} D_n(\omega;\omega_n) + \sum_{k=1}^{5} D_k^*(\omega;\omega_k), \quad (18)$$

where $n$ denotes the ionization shells and $k$ the excitation levels, and Drude type functions are given as:

$$D_n(\omega;\omega_n) = f_n \frac{\omega_p^2 \gamma_n \omega}{(\omega^2 - \omega_n^2)^2 + (\gamma_n \omega)^2}; \quad (19)$$

$$D_k^*(\omega;\omega_k) = f_k \frac{2\omega_p^2 (\gamma_k \omega)^3}{\left[(\omega^2 - \omega_k^2)^2 + (\gamma_k \omega)^2\right]^2}, \tag{20}$$

where $\hbar\omega_p$ is the free-electron plasmon energy, and the coefficients $\omega_{n,k}$, $f_{n,k}$ and $\gamma_{n,k}$ represent, respectively, the oscillator energy, strength and damping. $D_k^*$ is more sharply peaked at $\omega = \omega_k$ than $D_n$ at $\omega = \omega_n$ and, therefore, it is more suitable for representing the discrete excitation levels in the absorption spectrum.

The coefficients of the Drude functions are determined by a fit to the optical data of Heller et al. [48] under the constraints of the *f*-sum rule,

$$\int_0^\infty \omega D_{n,k}^{(*)}(\omega;\omega_{n,k}) d\omega = \frac{\pi}{2} f_{n,k} \omega_p^2. \tag{21}$$

The real part of the dielectric function in the optical limit can then be computed from

$$\mathrm{Re}\{\varepsilon(\omega)\} = 1 + \sum_{n=1}^4 D_n^{KK}(\omega;\omega_n) + \sum_{k=1}^5 D_k^{KK}(\omega;\omega_k), \tag{22}$$

where

$$D_n^{KK}(\omega;\omega_n) = 1 + \frac{1}{\pi} P \int_{-\infty}^{+\infty} \frac{D_n(\omega';\omega_n)}{\omega' - \omega} d\omega' = \frac{f_n \omega_p^2 (\omega_n^2 - \omega^2)}{(\omega_n^2 - \omega^2)^2 + (\gamma_n \omega)^2} \tag{23}$$

and

$$D_k^{KK}(\omega;\omega_k) = 1 + \frac{1}{\pi} P \int_{-\infty}^{+\infty} \frac{D_k^*(\omega';\omega_k)}{\omega' - \omega} d\omega' = \frac{f_k \omega_p^2 (\omega_k^2 - \omega^2)\left[(\omega_k^2 - \omega^2)^2 + 3(\gamma_k \omega)^2\right]}{\left[(\omega_k^2 - \omega^2)^2 + (\gamma_k \omega)^2\right]^2} \tag{24}$$

represent, respectively, the Kramers-Kronig (KK) pairs of $D_n$ and $D_k^*$, and $P$ stands for Cauchy principal value.

In the $q \to 0$ limit, the Drude dielectric function becomes equivalent to the Lindhard dielectric function. $\varepsilon(\omega, q=0)$ can be extended to finite $q$ according to dispersion relation. Up to 2nd order in $q$, the Lindhard dielectric function yields

$$\omega_n(q) = \omega_n + a_{\mathrm{RPA}} \frac{q^2}{2m} \approx \omega_n + \frac{q^2}{2m}, \tag{25}$$

where the random-phase-approximation (RPA) dispersion coefficient $a_{RPA} \approx 1$ for liquid water. It follows that Eq. (25) has the correct limiting behavior at both small and large $q$, $\omega_n(q \to 0) = \omega_n$ and $\omega_n(q \to \infty) = q^2/2m$, while approximating the RPA dispersion relation at intermediate $q$ [43].

From the dielectric function which is dependent on energy and momentum, the differential inelastic scattering cross-section (in the Born approximation) for each ionization shell and excitation level of the water molecule can be calculated from

$$\frac{d\sigma_{(n,k)}}{d\omega} = \frac{\hbar}{\pi a_0 NE} \int_{q_-}^{q_+} \mathrm{Im}\left\{\frac{-1}{\varepsilon_{n,k}(\omega,q)}\right\} \frac{dq}{q}, \tag{26}$$

where $\sigma$ is in units of area/molecule and the molecular density $N = 0.3343 \times 10^{23}$ molecules/cm³ for liquid water. The total cross-section for each ionization shell and excitation level follows directly from Eq. (26):

$$\sigma_{(n,k)} = \int_0^{E_{max}} \frac{d\sigma_{(n,k)}}{d\omega} d\omega, \tag{27}$$

where $E_{max} = E$ for excitations and $E_{max}^{(n)} = (E + B_n)/2$ for ionizations, with $B_n$ the binding energy of the $n$th shell.

At low energies, the exchange and correlation effects play an important role; therefore, for electron energies <1 keV, the Born approximation formula is corrected by using classic Coulomb-field correction and the exchange correction terms [49] according to ICRU report [50]. Coulomb-field correction accounts for the potential energy gained by the incident electron in the field of the target molecule. The Coulomb corrected differential cross-section $\frac{d\sigma^j}{d\omega}(\hbar\omega, E')$ is calculated with energy $E' = E + B_j + U_j$ for ionizations, where $B_j$ is the $j$th shell binding energy and $U_j$ is the kinetic energy of the $j$th shell electron. Similarly, it is calculated with $E' = E + 2E_j$ for excitations, where $E_j$ is the $j$th excitation energy. The exchange term is given as:

$$\frac{d\sigma_{ex}^j}{d\omega}(\hbar\omega, E) = \frac{d\sigma^j}{d\omega}(E + \hbar\omega_j - \hbar\omega, E) - \left(1 - \sqrt{\frac{\hbar\omega_j}{E}}\right) \sqrt{\frac{d\sigma^j}{d\omega}(\hbar\omega, E) \frac{d\sigma^j}{d\omega}(E + \hbar\omega_j - \hbar\omega, E)}. \tag{28}$$

The total differential cross-section is then the sum of the Coulomb's term and the exchange term,

$$\frac{d\sigma^j}{d\omega}(\hbar\omega, E) = \frac{d\sigma^j}{d\omega}(\hbar\omega, E') + \frac{d\sigma^j_{ex}}{d\omega}(\hbar\omega, E). \tag{29}$$

For electrons with energies above 10 keV the relativistic correction is considered [51].

B. K-shell

As the kinetic energy of K-shell electrons is relatively high ($U \sim 800$ eV), and their orbiting speed is comparable with that of the projectile, the model proposed in ICRU report [52] is applied. The differential cross-section for the K-shell is then:

$$\frac{d\sigma_K}{d\omega} = n\frac{4\pi\hbar\alpha_0^2 R^2 N}{E+B+U}\left\{\left[\frac{1}{(\hbar\omega)^2} - \frac{1}{\hbar\omega(E-\hbar\omega+B)} + \frac{1}{(E-\hbar\omega+B)^2}\right] + \frac{4U}{3}\left[\frac{1}{(\hbar\omega)^3} + \frac{1}{(E-\hbar\omega+B)^3}\right]\right\}, \tag{30}$$

where $B \sim 539.7$ eV is the binding energy of the K-shell, $n = 2$ the number of shell electrons, $R$ the Rydberg constant, $U \sim 800$ eV the average kinetic energy of the shell electrons. For $E > 540$ eV the K-shell contribution cannot be neglected.

By means of the above model which distinguishes between valence and core processes the total differential inelastic scattering cross-section is the sum,

$$\frac{d\sigma}{d\omega} = \sum_{n,k}\frac{d\sigma_{(n,k)}}{d\omega} + \frac{d\sigma_K}{d\omega}. \tag{31}$$

At lower energies, correction functions established empirically for H$_2$O were applied, they read: $Y_{ion} = \left[1 - 1.05\exp(-0.0088E^{1.1})\right]$ for the continuum where $E$ is the kinetic energy, and $Y_{j,exc} = \left[1 - (\hbar\omega_j/E)^a\right]^b$ for the discrete where $a$ and $b$ are constants. The Born corrected differential inelastic scattering cross-section is then calculated from

$$\frac{d\sigma}{d\omega} = Y_{ion}\sum_{n=1}^{4}\frac{d\sigma_n}{d\omega} + \sum_{k=1}^{5}Y_{k,exc}\frac{d\sigma_k}{d\omega} + \frac{d\sigma_K}{d\omega}. \tag{32}$$

The IMFP relates to the inelastic cross-section via $\lambda_{in}^{-1} = N\sum\sigma$. In G4DNA the inelastic scattering cross-sections are applicable down to 7.4 eV (the minimum electronic excitation potential of water molecule) [40-41]. Like other track structure codes (NOREC, PARTRAC etc.), in G4DNA an electron is stopped when its energy is lowered down to this cutoff energy of 7.4 eV, and is considered as solvated or thermalized while its remaining energy is assumed to be locally deposited. This was considered as adequate for energy deposition studies [40-41]. Also, Eq. (18) allows non-zero values for $\text{Im}\{\varepsilon_{(n,k)}(\omega,q)\}$ for all positive values of $\omega$, i.e. even at sub-ionization and sub-excitation energies, which is clearly unphysical. A common strategy to overcome this problem, also adopted in the G4DNA existing model, is to "cut" the Drude functions at $B_n$ so that they vanish below the ionization thresholds [43-44].

Below the threshold energy electron attachment and dissociation reactions become important [40-41,52-53]. In G4DNA, dissociative electron attachment cross-sections are available at energies 4-13 eV and vibrational excitation cross-section data set is available for energy range 2-100 eV [54-55].

## IV. Simulation with G4DNA

For simulation with G4DNA, we used Geant4 ver.10.4 on Ubuntu environment for the simulation of SEY and electron backscattering coefficient for liquid water. All materials were defined according to their composition stated in the NIST materials database (built into Geant4). The accuracy of the geometry was verified by activating G4PVPlacement constructor to check for any overlapping volumes. The number of primary electrons used was $10^6$ which were incident orthogonal to the target plane. We designed our simulation in such a manner that we can use different G4DNA physics options; however, we constructed our own physics list so that we can use modified Emfietzoglou model for inelastic scattering with Uehara screened Rutherford elastic scattering model.

## V. Results and Discussion

1. Secondary Electron Yield

The SEY was calculated as the ratio of the number of emitted secondary electrons with energies less than 50 eV to the number of primary electrons; the number of $10^6$ incident electrons are used at each condition). Fig. 2 shows a comparison on the SEY vs. primary energy curve, $\delta(E_p)$, between the simulation results from the two codes and the experimental results [56-58]. The experimental data shows very large deviations particularly below 1 keV where only two dada sets are avilable; the deviation may be attribured to different experimental procedures adopted as well as to the physical forms of water used for experiments. For example, Hilleret used monolayers of amorphous solid water on Cu substrate [58], Suszcynsky measured SEY for water ice [57], whereas Thiel measured liquid water [56]. It can be seen that the two codes, CMC and G4DNA, give the very similar SEY curve shape; however, there is a significant difference in the absolute SEY values calculated from the two codes. The SEY calculated from G4DNA model is much lower as compared to that calculated from CMC. It can be seen that the simulation results with our CMC code are much closer to the experimental results of Hilleret et al. below 1 keV, and to all three experimental data sets above 1 keV.

In order to identify the origin of the lower SEY values from G4DNA we illustrate in Fig. 3 the normalized energy spectra of secondary electrons calculated by the two codes. Our CMC simulation presents a universal distribution curve shape as that of metals [29], which peaks at 1.5 eV above vacuum level and having FWHM~4.8 eV. But G4DNA yields an abnormal curve shape which is abruptly cutoff at 7.4 eV and, hence, there is no usual low energy secondary electron signals below 7.4 eV. This behavior is obviously attributed to the fact that in G4DNA electrons with energies lower than 7.4 eV by default are not tracked; when an electron reaches the highest energy domain of G4DNA one step thermalization model (7.4 eV), it is then automatically converted into a solvated electron and displaced from its original position. The fundamental reason of setting such a minimum electronic excitation potential for water molecule is due to the Heller's optical dielectric function data [48] employed in the semi-empirical inelastic model [42,45] of the dielectric function, where the ELF is vanishing for $\hbar\omega < 7.4$ eV and, hence, the inelastic scattering cross-section becomes zero when electron energy is below 7.4 eV. To verify this, we have further reduced the upper limit of thermalization model by setting the smaller cutoff energies as 5-2 eV to enable electron transport at lower energy region above the threshold value and then calculated the SEY by G4DNA. Fig. 4 shows the calculated secondary electron energy spectra for different thresholds. As predicated the energy distribution cutoff is lowered down correspondingly and the increased area under the curve contributes to increased SEY values. Fig. 5 shows that as we reduce the upper limit of the thermalization model the SEY increases. However, when we compare with the inset of Fig. 2 it is clearly that the absolute SEY values by G4DNA still differ largely with experimental data of Hilleret [58] except by setting threshold around 4 eV, let along the energy distribution curve in Fig. 4. Table 1 shows a comparison on the maximum SEY value, $\delta_{max}$, and the corresponding maximum energy, $E_{max}$, obtained from the two MC codes with that of experimental data. The maximum energy is the primary energy at which the SEY reaches the maximum. From the table, we can observe that there is a general agreement in values of $E_{max}$ between experiment [58], CMC and G4DNA; but for $\delta_{max}$ still there is a certain difference.

As explained the threshold of 7.4 eV is due to the use of Heller's optical dielectric function data [48]. In our CMC simulation the optical data are taken from Segelstein [31] where the ELF is vanishing between 1-6 eV but is significant below 0.45 eV as can be seen in Fig. 1, therefore the low energy electrons ($E < 7$ eV) still have large inelastic scattering cross-section for energy loss due to vibrational excitation processes to slow down electrons. In this way the CMC code does not impose any cutoff on electron energy and can simulate complete secondary electron energy spectrum in the full range of 0-50 eV.

2. Electron backscattering coefficient

Electron backscattering coefficient, $\eta$, has also been calculated as the ratio of the number of backscattered electrons with energies greater than 50 eV to the number incident electrons. Fig. 6 compares the two codes for the values of $\eta$ in the primary energy range of 50 eV-30 keV. The results show good agreement on the $\eta(E_p)$ curve with a negligible difference on vales. Unfortunately, there is no sufficient experimental data for verification; the only available experimental data set is that of Joy [3] measured at high energies of 15-30 keV for liquid water using QuantomixTM capsules.

## VI. Conclusions

In conclusion, we have compared our CMC code with G4DNA code for the calculation of SEY and electron backscattering coefficient for liquid water in the primary energy range of 50 eV-30 keV. For electron backscattering coefficient the two codes agree with each other quite well. For the SEY, both CMC and G4DNA calculations yield the same curve shape of SEY dependence on primary energy but the absolute values are quite different. G4DNA underestimates the SEY values when compared with several experimental data sets. In addition, the energy distribution of secondary electrons calculated by G4DNA is abnormal, showing an abrupt cut off at 7.4 eV. This is due to the use of Heller's optical dielectric function data in the G4DNA inelastic model where the ELF is vanishing for $\hbar\omega < 7.4$ eV. Therefore, a full optical data by covering the low energy loss range ($1.24 \times 10^{-7}$ - 5.9 eV of Segelstein) shold be used in G4DNA model.

## Acknowledgement

This work was supported by the National Natural Science Foundation of China (No. 11574289). We thank Drs. S. Tanuma and H. Yoshikawa for providing the optical data of water, and also to Dr. H.M. Li and the supercomputing center of USTC for the support of parallel computing.

# References

[1]. M. Schenk, M. Futing and R. Reichelt, Direct visualization of the dynamic behavior of a water meniscus by scanning electron microscopy, J. Appl. Phys. 84, 4880-4884 (1998).

[2]. S. Thiberge, O. Zik and E. Moses, An apparatus for imaging liquids, cells, and other wet samples in the scanning electron microscopy, Rev. Sci. Instrum.75, 2281-2289 (2004).

[3]. D.C. Joy and C.S. Joy, Scanning electron microscope imaging in liquids- some data on electron interactions in water, J. Microsc. 221, 84-88 (2006).

[4]. R. Shimizu and Z.J. Ding, Monte Carlo modeling of electron-solid interactions, Rep. Prog. Phys. 55, 487-531 (1992).

[5]. D.C. Joy, Monte Carlo Modeling for Electron Microscopy and Microanalysis, Oxford Univ. Press, 216 (1995).

[6]. Z.J. Ding and R. Shimizu, A Monte Carlo modeling of electron interaction with solids including cascade secondary electron production, Scanning 18, 92-113 (1996).

[7]. Z.J. Ding, X.D. Tang and R. Shimizu, Monte Carlo study of secondary electron emission, J. Appl. Phys. 89, 718-726 (2001).

[8]. Z.J. Ding, H.M. Li, X.D. Tang and R. Shimizu, Monte Carlo simulation of absolute secondary yield of Cu, Appl. Phys. A 78, 585-587 (2004).

[9]. Z.J. Ding, H.M. Li, K. Goto, Y.Z. Jiang and R. Shimizu, Energy spectra of backscattered electrons in Auger electron spectroscopy: Comparison of Monte Carlo simulation with experiment, J. Appl. Phys. 96, 4598-4606 (2004).

[10]. H.M. Li and Z.J. Ding, Monte Carlo simulation of secondary electron and backscattered electron images in scanning electron microscopy for specimen with complex geometric structure, Scanning 27, 254-267 (2005).

[11]. Y.G. Li, P. Zhang and Z.J. Ding, Monte Carlo simulation of CD-SEM images for linewidth and critical dimension metrology, Scanning 35, 127-139 (2013).

[12]. Y.B. Zou, S.F. Mao, B. Da and Z.J. Ding, Surface sensitivity of secondary electrons emitted from amorphous solids: Calculation of mean escape depth by a Monte Carlo method, J. Appl. Phys. 120, 235102 (2016).

[13]. C.G. Frase, D. Gnieser and H. Bosse, Model-based SEM for dimensional metrology tasks in semiconductor and mask industry, J. Appl. Phys. 42, 183001 (2009).

[14]. H. Nikjoo, S. Uehara, D. Emfietzoglou and F.A. Cucinotta, Track-structure codes in radiation research, Radiat. Meas. 41, 1052-1074 (2006).

[15]. GEANT4 collaboration, Geant4: A simulation toolkit, Nucl. Instrum. Meth. A 506, 250-303 (2003).

[16]. GEANT4 collaboration, Geant4 and its validation, Nucl. Phys. B 150, 44-49 (2006).

[17]. S. Incerti, G. Baldacchino, M. Bernal, R. Capra, C. Champion and Z. Francis, The Geant4-DNA project, Int. J. Model Simul. Sci. Comput. 1, 57-78 (2010).

[18]. M.A. Bernal, M.C. Bordage, J.M.C. Brown, M. Davidkova, E. Delage, Z. El Bitar, S.A. Enger, Z. Francis, S. Guatelli, V.N. Ivanchenko, M. Karamitros, I. Kyriakou, L. Maigne, S. Meylan, K. Murakami, S. Okada, H. Payno, Y. Perrot, I. Petrovic, Q.T. Pham, A. Ristic-Fira, T. Sasaki, V. Stepan, H.N. Tran, C. Villagrasa, S. Incerti, Track structure modeling in liquid water: A review of the Geant4-DNA very low energy extension of the Geant4 Monte Carlo simulation toolkit, Phys. Med. 31, 861-874 (2015).

[19]. GEANT4 collaboration, Recent developments in Geant4, Nucl. Instrum. Meth. A 835, 186-225 (2016).

[20]. C. Li, S.F. Mao, Y.B. Zou, Y.G. Li, P. Zhang, H.M. Li and Z.J. Ding, A Monte Carlo modeling on charging effect for structures with arbitrary geometries, J. Phys. D: Appl. Phys. 51, 165301 (2018).

[21]. N.F. Mott, The scattering of fast electrons by atomic nuclei, Proc. R. Soc. London A 124, 425 (1929).

[22]. Y. Yamazaki, Studies on electron scattering by mercury atoms and electron spin polarization detector, Ph.D Thesis (Osaka University, 1977).

[23]. J.P. Desclaux, A multiconfiguration relativistic Dirac-Fock program, Comput. Phys. Commun. 9, 31-45 (1975); Erratum, Comput. Phys. Commun. 13, 71 (1977).

[24]. J.B. Furness and I.E. McCarthy, Semiphenomenological optical model for electron scattering on atoms, J. Phys. B 6, 2280-2291 (1973).

[25]. F. Salvat, A. Jablonski and C.J. Powell, ELSEPA—Dirac partial-wave calculation of elastic scattering of electrons and positrons by atoms, positive ions and molecules, Comput. Phys. Commun. 165, 157-190 (2005).

[26]. L.H. Yang, J. Tóth, K. Tőkési, B. Da and Z.J. Ding, Calculation of electron inelastic mean free path of three transition metals from reflection electron energy loss spectroscopy spectrum measurement data, Eur. Phys. J. D (2019, in press).

[27]. L. Reimer, Scanning Electron Microscopy (2nd Ed.), Springer, 1998.

[28]. D.R. Penn, Electron mean-free-path calculations using a model dielectric function, Phys. Rev. B 35, 482 (1987).

[29]. S.F. Mao, Y.G. Li, R.G. Zeng and Z.J. Ding, Electron inelastic scattering and secondary electron emission calculated without the single pole approximation, J. Appl. Phys. 104, 114907 (2008).

[30]. Z.J. Ding and R. Shimizu, Inelastic collisions of kV electrons in solids, Surf. Sci. 222, 313-331 (1989).

[31]. D.J. Segelstein, The complex refractive index of water, MSc thesis (University of Missouri-Kansas City, 1981).

[32]. H. Hayashi, N. Watanabe, Y. Udagawa and C.-C. Kao, The complete optical spectrum of liquid water measured by inelastic x-ray scattering, Proc. Natl. Acad. Sci. U.S.A. 97, 6264-6266 (2000).

[33]. B.L. Henke, E.M. Gullikson and J.C. Davis, X-Ray interactions: Photo-absorption, scattering, transmission, and reflection at E=50-30,000 eV, Z=1-92, At. Data Nucl. Data Tables 54, 181-342 (1993).

[34]. T. Kai, A. Yokoya, M. Ukai and R. Watanabe, Cross sections, stopping powers, and energy loss rates for rotational and phonon excitation processes in liquid water by electron impact, Radiat. Phys. Chem. 108, 13-17 (2015).

[35]. M.S. Chung and T.E. Everhart, Role of plasmon decay in secondary electron emission in the nearly-free-electron metals. Application to aluminum, Phys. Rev. B 15, 4699 (1977).

[36]. A.L. McNamara, S. Guatelli, D.A. Prokopovich, M.I. Reinhard and A.B. Rosenfeld, A comparison of X-ray and proton beam low energy secondary electron track structures using the low energy models of Geant4, Int. J. Radiat. Biol. 88, 164-170 (2012).

[37]. S. Chauvie, Z. Francis, S. Guatelli, S. Incerti, B. Mascialino, P. Moretto, P. Nieminen and M.G. Pia, Geant4 physics processes for microdosimetry simulation: Design foundation and implementation of the first set of models, IEEE Trans. Nucl. Sci. 54, 2619-2628 (2007).

[38]. S. Incerti, A. Ivanchenko, M. Karamitros, A. Mantero, P. Moretto, H. N. Tran, B. Mascialino, C. Champion, V.N. Ivanchenko, M.A. Bernal, Z. Francis, C. Villagrasa, G. Baldacchin, P. Guèye, R. Capra, P. Nieminen and C. Zacharatou, Comparison of GEANT4 very low energy cross section models with experimental data in water, Med. Phys. 37, 4692-4708 (2010)

[39]. C. Villagrasa, Z. Francis and S. Incerti, Physical models implemented in the Geant4-DNA extension of the Geant-4 toolkit for calculating initial radiation damage at the molecular level, Radiat. Prot. Dosim. 143, 214-218 (2011).

[40]. D. Emfietzoglou, G. Papamichael, K. Kostarelos and M. Moscovitch, A Monte-Carlo track structure code for electron (10 eV-10 keV) and protons (0.3-10 MeV) in water: partitioning of energy and collision events, Phys. Med. Biol. 45, 3171-3194 (2000).

[41]. D. Emfietzoglou, K. Karava, G. Papamichael and M. Moscovitch, Monte Carlo simulation of the energy loss of low-energy electrons in liquid water, Phys. Med. Biol. 48, 2355 (2003).

[42]. D. Emfietzoglou and M. Moscovitch, Inelastic collision characteristics of electrons in liquid water, Nucl. Instrum. Meth. B 193, 71-78 (2002).

[43]. I. Kyriakou and S. Incerti, Technical Note: Improvements in Geant4 energy-loss model and the effect on low-energy electron transport in liquid water, Med. Phys. 42, 3870 (2015).

[44]. I. Kyriakou, M. Sefl, V. Nourry and S. Incerti, The impact of new Geant4-DNA cross section models on electron track structure simulations in liquid water, J. Appl. Phys. 119, 194902 (2016).

[45]. S. Uehara, H. Nikjoo and D.T. Gwdhead, Cross-sections for water vapour for the Monte Carlo electron track structure code from 10 eV to the MeV region, Phys. Med. Biol. 37, 1841-1858 (1992).

[46]. D. Emfietzoglou, Inelastic cross-sections for electron transport in liquid water: a comparison of dielectric models, Radiat. Phys. Chem. 66, 373-385 (2003).

[47]. D.J. Brenner and M. Zaider, A computationally convenient parameterization of experimental angular distributions of low energy electrons elastically scattered off water vapor, Phys. Med. Biol. 29, 443-447 (1983).

[48]. J.M. Heller, R.N. Hamm, R.D. Birkhoff and L.R. Painter, Collective oscillation in liquid water, J. Chem. Phys. 60, 3483-3486 (1974).

[49]. M. Dingfelder, D. Hantke, M. Inokuti and H.G. Paretzke, Electron inelastic-scattering cross sections in liquid water, Radiat. Phys. Chem. 53, 1-18 (1998).

[50]. International Commission on Radiation Units and Measurements, Stopping powers for electrons and positrons, ICRU Report 37, (Bethesda MD, ICRU Publications), (1996).

[51]. C. Bousis, D. Emfietzoglou, P. Hadjidoukas, H. Nikjoo and A. Pathak, Electron ionization cross-section calculations for liquid water at high impact energies, Nucl. Instrum. Meth. B 266, 1185-1192 (2008).

[52]. International Commission on Radiation Units and Measurements, Secondary Electron Spectra from Charged Particle Interactions, ICRU Report 55, (Bethesda MD, ICRU Publications, 1996).

[53]. M. Dingfelder, R.H. Ritchie, J.E. Turner, W. Friedland, H.G. Paretzke and R.N. Hamm, Comparisons of calculations with PARTRAC and NOREC: Transport of electrons in liquid water, Radiat. Res. 169, 584-594 (2008).

[54]. M. Michaud, A. Wen and L. Sanche, Cross sections for low-energy (1-100 eV) electron elastic and inelastic scattering in amorphous ice, Radiat. Res. 159, 3-22 (2003).

[55]. J. Meesungnoen and J.-P. Jay-Gerin, Low-energy electron penetration range in liquid water, Radiat. Res. 158, 657-660 (2002).

[56]. B.L. Thiel, D.J. Stokes and D. Phifer, Secondary electron yield of water, Microsc. Microanal. 5 (Suppl. 2), 282-283 (1999).

[57]. D.M. Suszcynsky and J.E. Borovsky, Secondary electron yields of solar system ices, J. Geophys. Res. 97, 2611-2619 (1992); J. Geophys. Res. 98, 7499 (1993).

[58]. V. Baglin, J. Bojko1, O. Gröbner, B. Henrist, N. Hilleret, C. Scheuerlein and M. Taborelli, The secondary electron yield of technical materials and its variation with surface treatments, Proc. EPAC, Vienna, Austria (2000).

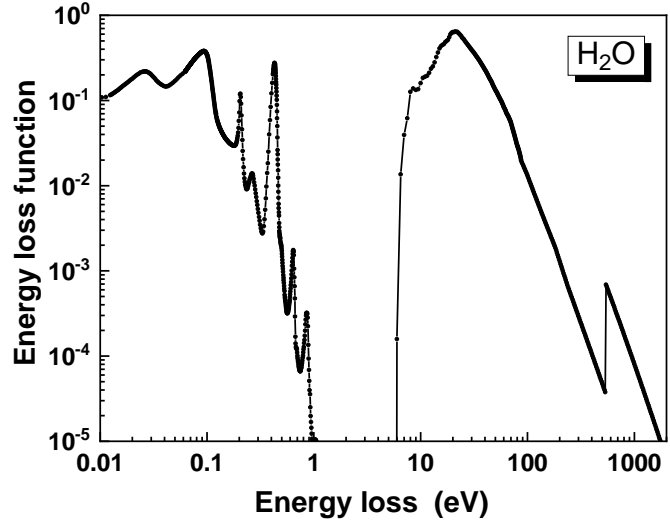

**Figure 1:** Optical ELF of liquid water as a function of energy loss or photon energy $\hbar\omega$ used in CMC simulation. The data were taken from Segelstein [31] in the photon energy range of $1.24\times10^{-7}$-5.9 eV, from Hayashi [32] in the range of 6-87 eV and from Henke [33] in the range of 87 eV-30 keV.

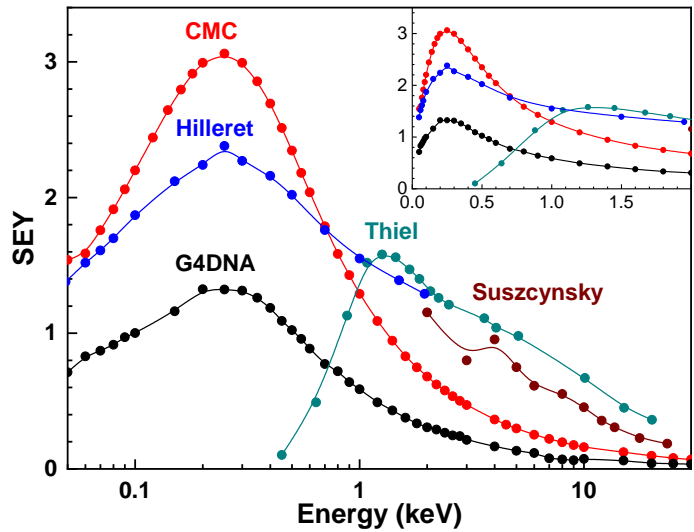

**Figure 2:** Comparison on the secondary electron yield as a function of incident electron energy between MC simulation results by the two codes and available experimental data of Thiel [56], Suszcensky [57] and Hilleret et al [58]. The inset is a linear plot in the low primary energy region.

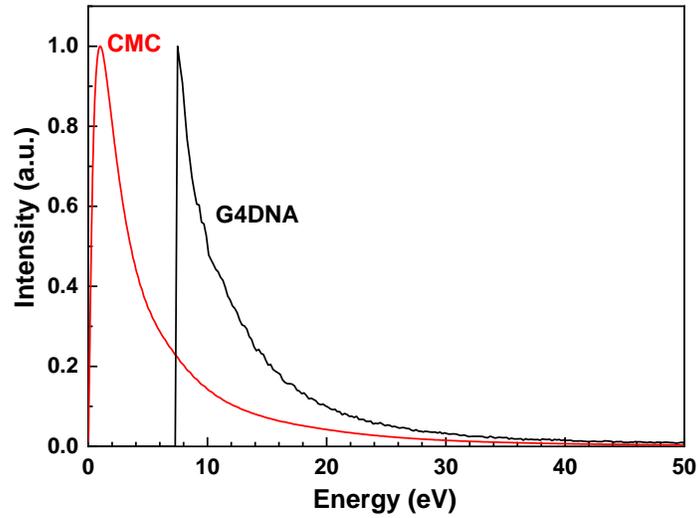

**Figure 3:** Comparison on the normalized secondary electron energy spectra simulated by using G4DNA and CMC at 1 keV incident energy.

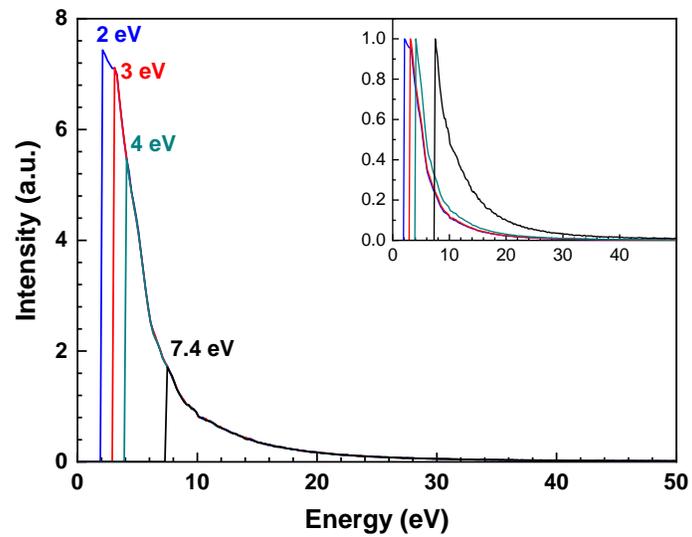

**Figure 4:** Simulated secondary electron energy spectra from G4DNA by setting different values of thermalization energy as 7.4 eV (default value), 4 eV, 3 eV and 2 eV; the inset is the corresponding normalized spectra.

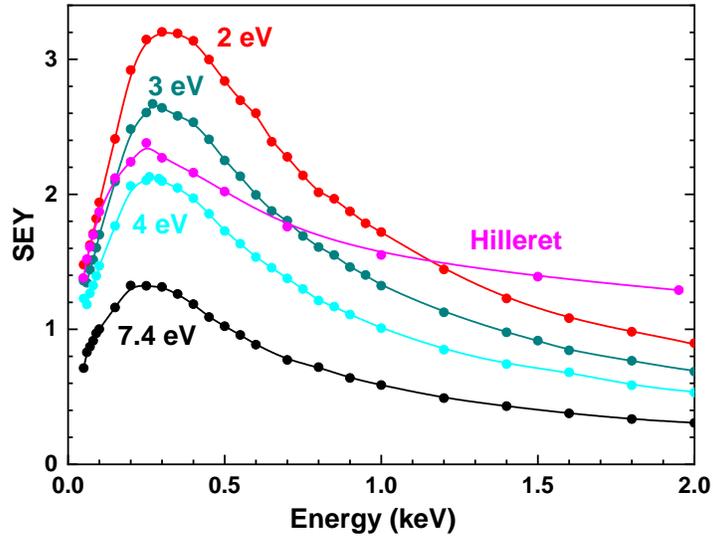

**Figure 5:** Simulated secondary electron yield from G4DNA by setting different values of thermalization energy as 7.4 eV (default value), 4 eV, 3 eV and 2 eV.

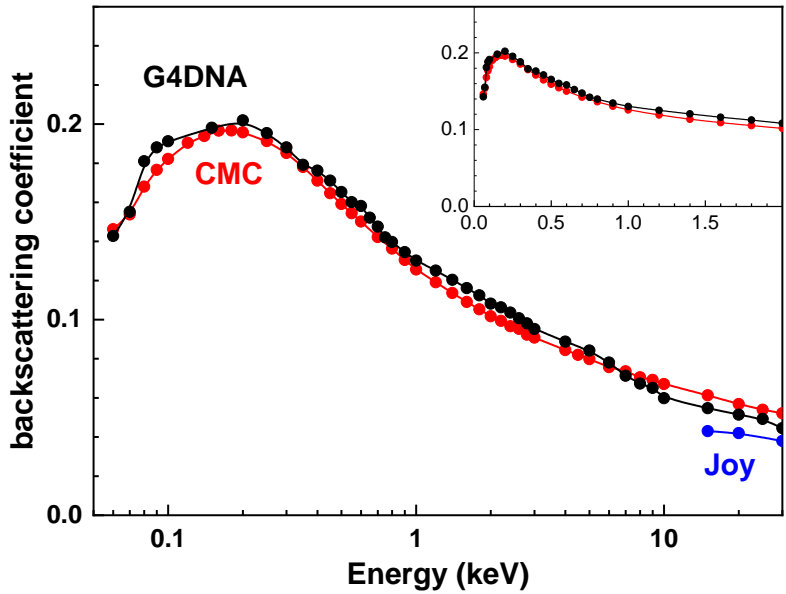

**Figure 6:** Comparison on electron backscattering coefficient as a function of incident electron energy between MC simulation results by the two codes and available experimental data.

**Table 1:** Comparison on the maximum SEY value and the corresponding maximum energy obtained from the two MC codes (where different cutoff values are employed for G4DNA) with that of experimental data.

|  | $E_{max}$ (eV) | $\delta_{max}$ |
|---|---|---|
| Exp. (Hilleret) | 250 | 2.4 |
| CMC | 250 | 3.06 |
| G4DNA (7.4 eV) | 200 | 1.32 |
| G4DNA (5 eV) | 250 | 1.70 |
| G4DNA (4 eV) | 260 | 2.13 |
| G4DNA (3 eV) | 270 | 2.67 |
| G4DNA (2 eV) | 300 | 3.20 |